# High-harmonic generation from subwavelength silicon films

K. Hallman[1], S. Stengel[2], W. Jaffray[2], F. Belli[2], M. Ferrera[2], M.A. Vincenti[3], D. de Ceglia[3], Y. Kivshar[4], N. Akozbek[5], S. Mukhopadhyay[6], J. Trull[6], C. Cojocaru[6], and M. Scalora[7*]

[1]*PeopleTec, Inc. 4901-I Corporate Dr., Huntsville, AL 35805, USA*
[2]*Institute of Photonics and Quantum Sciences Heriot-Watt University, SUPA Edinburgh, EH14 4AS United Kingdom*
[3]*Department of Information Engineering – University of Brescia, 25123 Brescia, Italy*
[4]*Nonlinear Physics Centre, Australian National University, Canberra, ACT 2601, Australia*
[5]*US Army Space & Missile Defense Command, Tech Center, Redstone Arsenal, AL 35898 USA*
[6]*Department of Physics, Universitat Politècnica de Catalunya, 08222 Terrassa (Barcelona), Spain*
[7]*FCDD-AMT-MGR, DEVCOM AvMC, Charles M. Bowden Research Center, Redstone Arsenal, Alabama, 35898-5000, USA*

*\*michael.scalora.civ@army.mil*

**Abstract**

Recent years have witnessed significant developments in the study of nonlinear properties of various optical materials at the nanoscale. However, in most cases experimental results on harmonic generation from nanostructured materials are reported without the benefit of suitable theoretical models and appropriate comparisons to assess enhancement of conversion efficiencies compared to the intrinsic properties of a given material. Here, we report experimental observations of *even and odd optical harmonics* generated from a suspended subwavelength silicon film, a dielectric membrane, up to the 7$^{th}$ harmonic tuned deep in the UV range at 210nm, which is the current limit of our detection system, using peak power densities of order 3TW/cm$^2$. We explain the experimental data with a time domain, hydrodynamic-Maxwell approach broadly applicable to most materials. Our approach accounts simultaneously for surface and magnetic nonlinearities that generate *even optical harmonics*, as well as linear and nonlinear material dispersions beyond the third order to account for *odd optical harmonics*, plasma formation, and a phase locking mechanism that makes the generation of high harmonics possible deep into the UV range, where semiconductors like silicon start operating in a metallic regime.

The last decade has witnessed a renewed quest for efficient nonlinear frequency conversion implementing micrometric and nanometric size materials and devices that have highlighted the importance of nanostructures and metasurfaces. This pursuit has been marked by a dramatic increase in theoretical and experimental work [1-12] that has led to a natural reassessment of the optical properties of semiconductors and metals in regimes that were previously deemed impractical or inaccessible [13-21]. This kind of research has generally been limited to second and third harmonic generation (SHG and THG, respectively) primarily due the applicability of these processes to surface sensing, label-free bio-imaging, and quantum optics [22-25]. However, nonlinear high harmonic generation (HHG) encompassing visible light, ultraviolet (UV), and even shorter wavelengths, is also of fundamental interest for other key applications such as ultra-fast physics, and X-UV photolithography [26, 27]. Most work on HHG has been carried out in transparent conductive oxides (TCOs) [28-31] and all dielectric structures [5-6] for the purpose of avoiding intrinsic losses associated with semiconductors and metals [32]. This has led researchers to not only set metals aside but also to focus almost exclusively on the transparency range of dielectric and semiconductor materials.

Generally, the study of nonlinear optics at the nanoscale requires modifications of the material equations of motion to account for effects that manifest themselves only at the atomic level, given that the average atomic diameter is of order 3Å-5Å. Examples are linear and nonlinear nonlocal effects due to pressure and viscosity of the free electron fluid and spill-out effects [33-37], as well as tunneling [38, 39], screening and surface and magnetic phenomena [40-42]. Even in transparent materials, most of the work has focused on bulk nonlinearities, while neglecting crucial surface and magnetic phenomena that in centrosymmetric materials have been traditionally associated with the generation of even-order harmonics [40]. Most semiconductors have broad absorption resonances deep in the UV range, perhaps suggesting that absorption may not be circumvented, with negative dielectric constants and epsilon near zero (ENZ) crossing points in the 100nm range. However, it has been demonstrated that SHG and THG can occur despite absorption in the visible and UV ranges, and can be transmitted through half millimeter thick wafers of silicon (Si), gallium arsenide (GaAs), and gallium phosphide (GaP) [13-16]. Although for nanometer-thick materials the conversion process is inefficient due to the absence and loss of meaning of phase matching, efficiency can be improved dramatically if a resonant mechanism comes into play as demonstrated for semiconductor cavities [17], and at least in theory in GaAs-filled metal gratings [18] and silicon nanowire arrays [19]. This phenomenon plays a critical role in nonlinear optical phenomena for harmonic generation in all semiconductors in the opacity range, and so we now briefly delve into the fundamental reasons that make this possible.

As originally pointed out by Bloembergen and Pershan [43] for SHG in transparent media, Maxwell's equations in a nonlinear material have two solutions: a homogeneous component that propagates with a wave vector $k_{2\omega}^{(hom)} = n(2\omega)2\omega/c$, and an inhomogeneous component having wave vector $k_{2\omega}^{(inh)} = 2k_\omega = 2n(\omega)\omega/c$. These distinct solutions can be generalized to the $j^{th}$ harmonic. In the plane wave regime and oblique incidence, these waves refract at different angles, interfere, and give rise to Maker fringes [44], with the inhomogeneous portion propagating at the same refraction angle and velocity as the pump beam, a phenomenon more recently clearly observed in bulk lithium niobate [45]. However, if the interaction occurs in such a way that the pump is tuned in the transparency range [$n(\omega)$ is real], and the harmonic field is tuned in the opaque region [$n(2\omega)$ is complex], then the homogenous component is absorbed, leaving behind the inhomogeneous component, setting up an anomalous situation, i.e., a harmonic signal that propagates with the same dispersion and group velocity of the pump, a phenomenon that has been referred to as phase locking, given the phase relationships between pump and the relevant inhomogeneous harmonic solutions [13-20, 45].

Thanks to phase locking, a harmonic field tuned in the opaque regime can resonate along with the pump, as if absorption were not present for that component, provided the pump is tuned in the transparency range. This is the fundamental mechanism that explains the apparent suppression of absorption observed for example, in references [8] and [9], and most recently highlighted and reported in the opacity range of a chalcogenide glass grating at 354nm [46], well inside the band gap, and in silicon etalons [47], with THG conversion efficiencies of order $10^{-7}$ at 266nm, with a pump intensity of 1.5GW/cm$^2$. The remarkable facts here are that one is exploiting the intrinsic properties of silicon without local field amplification [see Supplemental Information], and that silicon is metallic [Re($\varepsilon$)<0] between 100nm and 300nm. This efficiency should be contrasted with purely experimental observations of THG from three-dimensional (3-D), resonant silicon metasurfaces characterized by guided mode resonances and bound states in the continuum [48, 49], where similar peak power densities yield THG conversion efficiencies of order $10^{-4}$. In the present case, using the same theory illustrated in references [13], [19], and [47], we analyze experimental results that yield high harmonic generation in silicon using a time-domain, hydrodynamic-Maxwell pulse propagation method that considers linear and nonlinear bulk material dispersions, nonlinearities triggered by the mere traversal of surfaces, and by the intrinsically nonlinear magnetic portion of the Lorentz force. As we will see below, the confluence of the above-listed physical phenomena enables an accurate theoretical description of the relevant qualitative and quantitative aspects of nonlinear frequency conversion in the perturbative regime. We note that it is possible to extend the method to the non-perturbative regime by adopting additional polarization and inversion components generated by a system of Bloch equations that describe a two- or multi-level atom, as shown in references [50] and [51], where it was applied to stimulated Raman scattering near plasmonic structures.

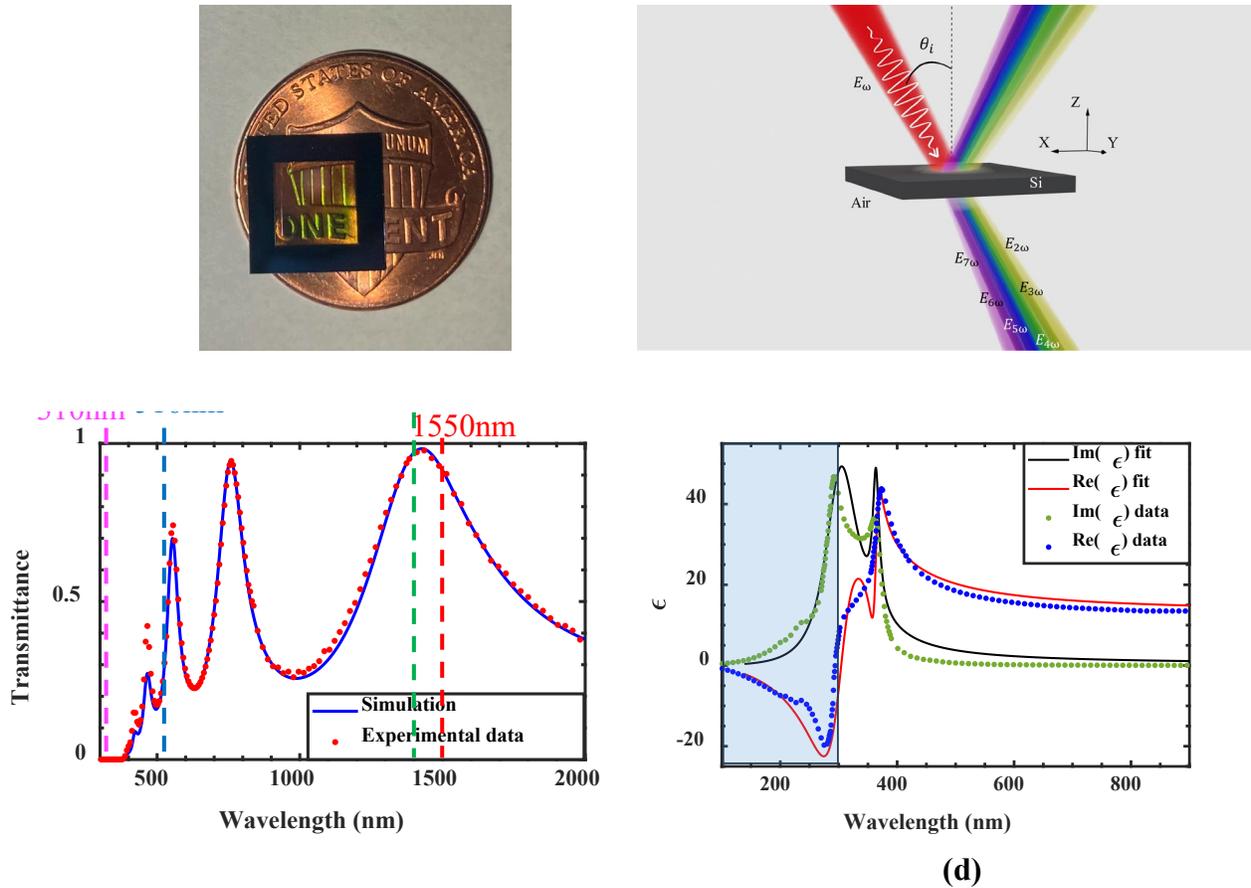

**Fig.1:** (a) Picture of a 200nm-thick silicon film. The surface area of the suspended portion of the film (the area with greenish hue) is approximately 5mm x 5mm. (b) A pulse is incident from the top. We monitor only transmitted fields. (c) Linear transmittance measured at normal incidence (red markers) and fitted (blue curve) using Palik's data in reference [32]. Some wavelengths of interest are marked in the figure. (d) Dielectric function of silicon as reported in reference [32] (markers) and as fitted by two Lorentzian functions (solid curves.) The dielectric function displays two resonances in the 300-400nm range, with rapid variations below 500nm. Silicon is metallic below 300nm (shaded region).

The suspended sub-wavelength silicon films used in our experiments are pictured and schematically depicted in Fig.1(a) and (b). The resonance near 530nm gives the sample a characteristic green hue. They were purchased from Norcada (Alberta, Canada) and consist of ~200nm-thick <100> silicon film with doping level of order $10^{16}/cm^3$, etched out of a silicon carrier wafer, and held by a frame of proprietary composition. This density corresponds to a plasma frequency in the THz range and does not affect the dynamics at low intensities. We will return to the subject later. In Fig.1(c) we report both measured and fitted transmittance at normal incidence. The etalon displays a series of Fabry-Perot resonances whose amplitude expectedly decreases rapidly with decreasing wavelength. Our measurements are well represented by Palik's data [32], which we show in Fig.1(d) along with a fit obtained using two detuned Lorentzian functions that we will use in our simulations. The sample is completely opaque below 400nm, erroneously suggesting that we should not expect any meaningful nonlinear response in this range.

Two experimental setups were used, one to cover power densities below 100GW/cm², and the other to cover the range between 200GW/cm² and 3TW/cm². They are both described in the Supplement. The dispersion of undoped silicon is marked by two adjacent resonances in the UV range [13, 32, 47], which according to Miller's rule [52, 53] give rise to resonant nonlinearities in the opaque range. The system may then be described by two polarization components having two closely spaced resonance frequencies and separate linear and nonlinear spring constants, as follows:

$$\ddot{\mathbf{P}}_{bj}^{(\tau)} + \tilde{\gamma}_{bj}\dot{\mathbf{P}}_{bj}^{(\tau)} + \tilde{\omega}_{0,bj}^2 \mathbf{P}_{bj} - \tilde{\beta}_{bj}(\mathbf{P}_{bj}\bullet\mathbf{P}_{bj})\mathbf{P}_{bj} + \tilde{\delta}_{bj}(\mathbf{P}_{bj}\bullet\mathbf{P}_{bj})^2\mathbf{P}_{bj} - \tilde{\theta}_{bj}(\mathbf{P}_{bj}\bullet\mathbf{P}_{bj})^3\mathbf{P}_{bj}$$
$$= \pi\tilde{\omega}_{pj}^2\mathbf{E} + \frac{e\lambda_r}{m_{bj}^*c^2}(\mathbf{P}_{bj}\bullet\nabla_{\xi,\varsigma,\zeta})\mathbf{E} + \frac{e\lambda_r}{m_{bj}^*c^2}\dot{\mathbf{P}}_{bj}^{(\tau)}\times\mathbf{H} \quad (1)$$

Here, $m_{bj}^*$ is the effective bound electron mass for the $j^{th}$ polarization component; $\tilde{\omega}_{0,bj}$ and $\tilde{\omega}_{pj}$ are the scaled resonance and plasma frequencies, respectively; $\lambda_r = 1\mu m$ is a reference wavelength that scales $\xi = z/\lambda_r$, $\zeta = x/\lambda_r$, $\varsigma = y/\lambda_r$, where x and y are the transverse coordinates, and z is the longitudinal coordinate; $\tau = ct/\lambda_r$ is the scaled time, and $c$ is the speed of light in vacuum. Spatial and temporal derivatives on the fields are performed with respect to these new scaled spatial and temporal coordinates, as represented by the superscript $(\tau)$ and the subscripts $(\xi,\varsigma,\zeta)$.

We now outline a novel approach to solving Eqs.1. The parameters $\tilde{\beta}_{bj} \approx \omega_{0,bj}^2 \lambda_r^2 / \left(L^2 n_{0,bj}^2 e^2 c^2\right)$, $\tilde{\delta}_{bj} = \tilde{\beta}_{bj} / \left(L^2 n_{0,bj}^2 e^2\right)$, and $\tilde{\theta}_{bj} = \tilde{\beta}_{bj} / \left(L^4 n_{0,bj}^4 e^4\right)$ are higher order scaled coefficients in a perturbative expansion that are derived directly from a nonlinear classical oscillator model, and are usually assumed to be constant. $n_{0,bj}$ is the bound electron density, and $L$ is usually interpreted as the lattice constant. From the point of view of a classical spring and classical macroscopic electrodynamics, the lattice constant $L$ represents the spring's maximum allowed extension, which is generally taken to be constant across the entire spectral range. From an atomic point of view, $L$ may be interpreted as either an atomic/orbital diameter or inter-particle distance, which for solids can vary from a fraction of 1Å to several Ås, a disparity that will be reflected in the particle density, and that will substantially affect the magnitudes of $\tilde{\beta}$, $\tilde{\delta}$, and $\tilde{\theta}$. *However, since harmonic generation occurs over a wide wavelength range, all the parameters outlined above are unlikely to remain constant, including electronic effective masses, due to the different energy levels that are found deeper and deeper inside the atom, with complex and primarily unknown orbital radii, population density, and band curvatures.* Therefore, orbitals and energy levels that are excited at the pump wavelength are different compared to states excited at each harmonic wavelength, will have different diameters and effective particle densities depending on the number of electrons in each orbital, so that at least the parameters $L$ and $n_0$ at the pump wavelength may be different at each of the harmonics. From a classical, macroscopic point of view, retrieval of $\tilde{\beta}$, $\tilde{\delta}$, and $\tilde{\theta}$ should be done in the same manner that the dielectric constant $\varepsilon$ is retrieved, i.e. via ellipsometric techniques for a given thickness and wavelength. Then, it is reasonable to expect that $\varepsilon$, $\tilde{\beta}$, $\tilde{\delta}$, and $\tilde{\theta}$ will display dispersive behavior not only as a function of wavelength, but also as a function of geometrical dispersion (resonances) [54]. Therefore, while it is generally acceptable to assume that $\tilde{\beta}$ is constant under most circumstances [47, 52, 53], we may well have $\tilde{\beta} = \tilde{\beta}_0 \, f_{\tilde{\beta}}(\lambda)$. Here, $\tilde{\beta}_0 = \omega_{0,bj}^2 \lambda_r^2 / \left(L_0^2 n_{0,bj}^2 e^2 c^2\right)$ is defined using the known lattice constant and particle density for the material, and $f_{\tilde{\beta}}(\lambda)$ is a parameter that generalizes the third order nonlinear coefficient in Eqs.1 that reflects the changing nature of the material across the spectrum. By the same token, we will assume that $\tilde{\delta} = \tilde{\delta}_0 \, f_{\tilde{\delta}}(\lambda)$ and $\tilde{\theta} = \tilde{\theta}_0 \, f_{\tilde{\theta}}(\lambda)$, with $\tilde{\delta}_0 = \tilde{\beta}_0 / \left(L_0^2 n_{0,bj}^2 e^2\right)$ and $\tilde{\theta}_0 = \tilde{\beta}_0 / (L_0^4 n_{0,bj}^4 e^4)$. Finally, given the proximity of the material resonances in Fig.1d, for simplicity we will assume that the coefficients $\tilde{\beta}_{bj}$, $\tilde{\delta}_{bj}$, and $\tilde{\theta}_{bj}$ have similar amplitudes at both resonance wavelengths.

Since we consider up to 7$^{th}$ harmonic generation, the bulk nonlinear response from bound electrons must reflect at least a 7$^{th}$ order nonlinearity, and may be written as:

$$\mathbf{P}_{NL,j} = -\tilde{\beta}_{bj}(\mathbf{P}_{bj} \bullet \mathbf{P}_{bj})\mathbf{P}_{bj} + \tilde{\delta}_{bj}(\mathbf{P}_{bj} \bullet \mathbf{P}_{bj})^2 \mathbf{P}_{bj} - \tilde{\theta}_{bj}(\mathbf{P}_{bj} \bullet \mathbf{P}_{bj})^3 \mathbf{P}_{bj} \quad . \quad (2)$$

Material response is assumed to be isotropic but could easily be modified to account for anisotropies [55] by isolating the spatial coordinates and by assigning the coefficients different values in different directions. In reference [55], which to our knowledge is the first report of reflected THG from silicon in the UV range, the authors investigated the question of THG using a nanosecond pump tuned to 1064nm and found an effective anisotropy in the third order nonlinear response. In the low intensity regime (100GW/cm$^2$ or less) integration of Eqs.(1) together with Maxwell's equations account for linear and nonlinear dispersions, and allow us to describe the process by considering the consequences of terms like $\frac{e\lambda_r}{m_{0,bj}^* c^2}(\mathbf{P}_{bj} \bullet \nabla_{\xi,\varsigma,\zeta})\mathbf{E}$, which represents surface nonlinearities, and the magnetic Lorentz contribution $\frac{e\lambda_r}{m_{0,bj}^* c^2}\dot{\mathbf{P}}_{bj}^{(\tau)} \times \mathbf{H}$, which contains both surface and volume nonlinear bound currents. Both terms are also expanded up to their 7$^{th}$ harmonic contributions (not shown here), and partially account for even harmonic generation in centrosymmetric materials that act as insulators. As we will see below, observation of the 7$^{th}$ harmonic at 210nm using our detection system will require peak power densities of order 3TW/cm$^2$. At these power densities and 10Hz repetition rate our incident beam delivers an energy density of approximately 270mJ/cm$^2$. Nevertheless, we expect single-pulse effects given the temporal displacement between consecutive pulses, each containing approximately 27mJ/cm$^2$. The threshold for plasma formation in 3μm-thick Si layers has been reported to be of order 150mJ/cm$^2$ when it is delivered within ~200fs using optimized pulses and yielding remarkably high free electron densities of order 10$^{22}$-10$^{23}$/cm$^3$ [56]. In our case, our calculations suggest free carrier densities of order 10$^{19}$-10$^{20}$/cm$^3$. Plasma formation requires an additional, transient polarization component $\mathbf{P}_f$ to include a description of the dynamics of free carriers. This component triggers additional effects like pump shielding and an intensity dependent plasma frequency (increasing carrier density) that tends to blueshift as a function of incident peak power density. In turn this lowers the effective dielectric constant dynamically, affecting the structure and position of the geometrical resonances, including modifying the coefficients $\tilde{\beta}$, $\tilde{\delta}$, and $\tilde{\theta}$, which continue to be dispersive and have to be modified accordingly. Put another way, the increased free charge density comes at the expense of a decreasing bound charge density, causing the coefficients $\tilde{\beta}$, $\tilde{\delta}$, and $\tilde{\theta}$ to change dynamically as a redistribution of charge occurs. Given these factors, the essential equation of motion that supplements Eq.1 and describes the dynamics of free carriers may be written as follows [13, 41, 42]:

$$\ddot{\mathbf{P}}_f^{(\tau)} + \tilde{\gamma}_f \dot{\mathbf{P}}_f^{(\tau)} = \left(\pi\tilde{\omega}_{p,f}^2 + \kappa_f \mathbf{E}\cdot\mathbf{E}\right)\mathbf{E} + \frac{e\lambda_r}{m_f^* c^2}\left(\nabla_{\xi,\varsigma,\zeta}\cdot\mathbf{P}_f\right)\mathbf{E} + \frac{3}{5}\frac{E_F}{m_f^* c^2}\left[\nabla_{\xi,\varsigma,\zeta}\left(\nabla_{\xi,\varsigma,\zeta}\cdot\mathbf{P}_f\right) + \frac{1}{2}\nabla_{\xi,\varsigma,\zeta}^2\mathbf{P}_f\right]$$

$$+\frac{e\lambda_r}{m_f^* c^2}\dot{\mathbf{P}}_f^{(\tau)}\times\mathbf{H} - \frac{1}{n_{0f}e\lambda_r}\left[\left(\nabla_{\xi,\varsigma,\zeta}\cdot\dot{\mathbf{P}}_f^{(\tau)}\right)\dot{\mathbf{P}}_f^{(\tau)} + \left(\dot{\mathbf{P}}_f^{(\tau)}\cdot\nabla_{\xi,\varsigma,\zeta}\right)\dot{\mathbf{P}}_f^{(\tau)}\right]$$

. (3)

We will not dwell on the Eq.3 other than to mention that the coefficient $\kappa_f$, which may depend on fluence and may also be dispersive, allows the scaled plasma frequency $\tilde{\omega}_{p,f}$ to change as a function of intensity (blueshift) and time, with contributions to *all* the harmonics, as revealed by an expansion of the fields up to 7$^{th}$ harmonic. The term proportional to the Fermi energy, $E_F = \frac{\hbar^2}{2m_f^*}\left(3\pi^2 n_{0f}\right)^{2/3}$, contains the effects of linear pressure and viscosity, $m_f^*$ and $n_{0f}$ are the free electron mass and background density, respectively, and $\tilde{\gamma}_f$ is the scaled damping coefficient. The terms in Eq.3 thus reflect a continuity equation that yields contributions to an *effective plasma frequency* that depends on the local charge density $\nabla_{\xi,\varsigma,\zeta}\cdot\mathbf{P}_f$ and a third order nonlinearity as follows [57]:

$$n(\mathbf{P},\mathbf{E}) = \left(n_{0f} - \frac{1}{e}\nabla_{\xi,\varsigma,\zeta}\cdot\mathbf{P}_f + \kappa_f \mathbf{E}\cdot\mathbf{E} + ...\right) \quad (4).$$

Nonlocal terms $\left(\nabla_{\xi,\varsigma,\zeta}\left(\nabla_{\xi,\varsigma,\zeta}\cdot\mathbf{P}_f\right) + \frac{1}{2}\nabla_{\xi,\varsigma,\zeta}^2\mathbf{P}_f\right)$ then make the dielectric function a highly dynamic variable.

In Fig.2 we show the measured and simulated spectra generated by a ~3TW/cm$^2$, 85fs pump pulse tuned to 1475nm incident normal to the surface to temporarily suppress even harmonics. The measured conversion efficiencies are 4.2x10$^{-5}$, 6.2x10$^{-7}$, and 4.57x10$^{-9}$ for 3$^{rd}$, 5$^{th}$, and 7$^{th}$ harmonic, with peak power densities of order 126MW/cm$^2$, 1.8MW/cm$^2$ and 14kW/cm$^2$, respectively, despite being tuned in the absorption and metallic ranges. Agreement between the measured and simulated data is remarkable. Phase-locking is the fundamental mechanism that makes this kind of interaction possible. Our simulations suggests that results like those depicted in Fig.2 can be obtained by tuning the pump field at each of the resonances shown in Fig.1c, pushing the highest harmonic we can presently simulate below 100nm. See Supplemental Information for a scan of 7$^{th}$ harmonic generation efficiency for a range of wavelength that includes the resonances shown in Fig.1c.

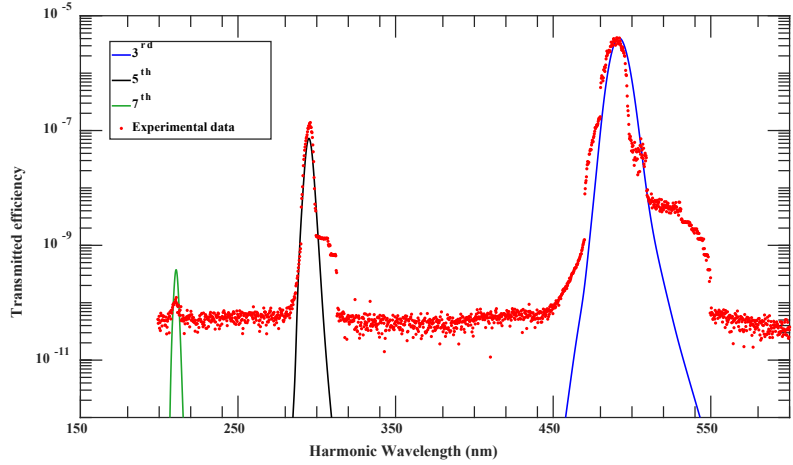

**Fig.2**: Measured and simulated spectra of transmitted 3$^{rd}$, 5$^{th}$, and 7$^{th}$ harmonic conversion efficiencies for a single pulse 85fs in duration, having carrier wavelength tuned to 1475nm, peak power density of 3TW/cm$^2$, and normally incident on the silicon film. Relevant parameters: $\tilde{\beta}(\omega)=5\times10^{-10}$, $\tilde{\beta}(3\omega)=2\times10^{-10}$ $\tilde{\beta}(5\omega)=5\times10^{-11}$ $\tilde{\beta}(7\omega)=5\times10^{-11}$, $\tilde{\delta}(\omega)=1.25\times10^{-19}$, $\tilde{\delta}(3\omega)=5\times10^{-20}$, $\tilde{\delta}(5\omega)=1.25\times10^{-20}$ $\tilde{\delta}(7\omega)=1.25\times10^{-20}$, $\tilde{\vartheta}(\omega)=6.25\times10^{-29}$, $\tilde{\vartheta}(3\omega)=2.5\times10^{-29}$ $\tilde{\vartheta}(5\omega)=6.25\times10^{-30}$ $\tilde{\vartheta}(7\omega)=6.25\times10^{-30}$, $k_f(\omega)=1.6\times10^{-9}$, $k_f(3\omega)=6.4\times10^{-10}$, $k_f(5\omega)=1.6\times10^{-10}$, $k_f(7\omega)=1.6\times10^{-10}$.

We then pump the sample at 1550nm at normal incidence and perform an intensity scan to detect *third* and *fifth* harmonic signals at 516nm and 310nm, which is our detection limit for the low-intensity setup. In Fig.3(a-b) we show the result of both scans, along with the result of our simulations. Conversion efficiency is defined by dividing the total energy of a given harmonic by the total incident pump energy. Once the dielectric constant is established and fitted using the linearized version of Eqs.1 [13], $\tilde{\beta}$ and $\tilde{\delta}$ are the only remaining parameters to be determined.

The reconstruction of nonlinear dispersion when $\tilde{\delta}$ is not zero is complicated because the $\tilde{\delta}$ terms tends to quench the $\tilde{\beta}$ contributions at both the pump and its harmonic wavelengths, given their opposite signs. This effect is discussed in reference [58]. The detection system in Fig.S1(a) can be used to detect signals down to approximately 310nm. Therefore, a 7[th] harmonic was also observed when the pump pulse was tuned to a carrier wavelength of 2150nm. We show the measured 7[th] harmonic efficiency vs. incident peak power density in Fig.3(c). We note that the transmission function in Fig.1(c) expresses the behavior of a *pump field* tuned to that wavelength and says nothing about the behavior of a harmonic signal generated and tuned to the same wavelength. This conclusion is fully supported by the fact that the generated 5[th] harmonic signal is transmitted by the etalon, notwithstanding the fact that the linear transmission function suggests that wavelength should be fully suppressed. We emphasize that transmission of the 210nm and 310nm signals occurs because they are the inhomogeneous components of the generated harmonics, which propagate under phase locking conditions, i.e., with the dispersive properties of the pump field tuned in the transparency range of the material.

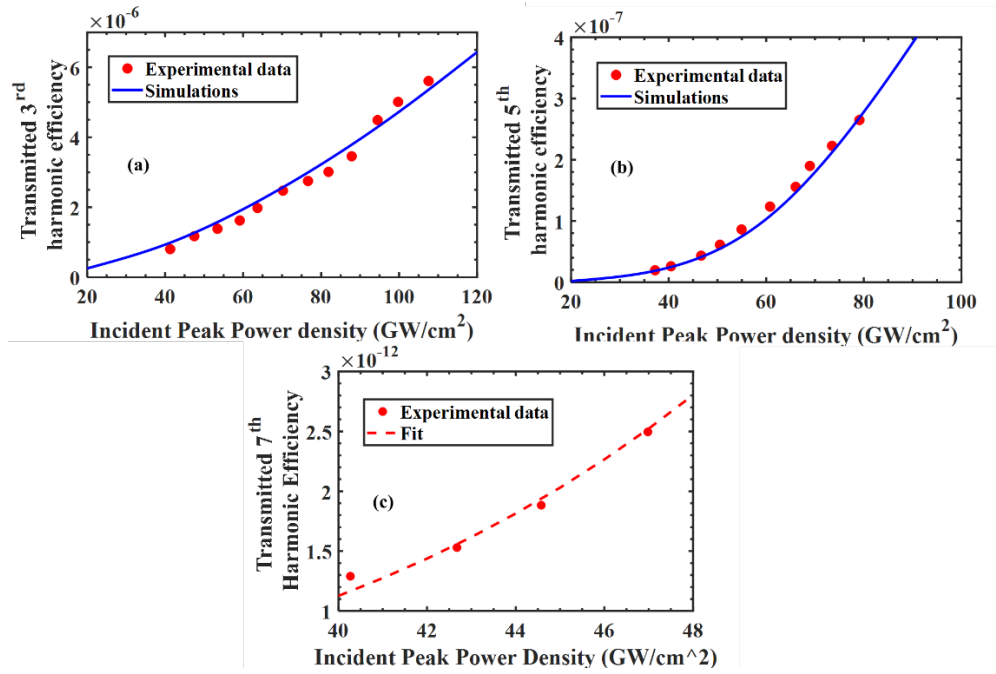

**Fig.3**: Transmitted third (a) and fifth (b) harmonic generation efficiency vs incident peak power density. An intensity scan is performed with a ~100fs pulses tuned to 1550nm with repetition rate 2Hz. The 3rd (516nm) and 5th (310nm) harmonic data are fitted almost perfectly by the simulations. Relevant parameters: $\tilde{\beta}(3\omega) = \tilde{\beta}_0 = 8.5 \times 10^{-9}$, $\tilde{\delta}(3\omega) = \tilde{\delta}_0 = 4.25 \times 10^{-18}$, $\tilde{\vartheta}(3\omega) = \tilde{\vartheta}_0 = 2.125 \times 10^{-27}$, $\tilde{\beta}(5\omega) = \tilde{\beta}_0$, $\tilde{\delta}(5\omega) = 9\tilde{\delta}_0$, $\tilde{\vartheta}(5\omega) = 5\tilde{\vartheta}_0$, $\tilde{\beta}(7\omega) = 4\tilde{\beta}_0$, $\tilde{\delta}(7\omega) = \tilde{\delta}_0$, $\tilde{\vartheta}(7\omega) = \tilde{\vartheta}_0$. (c) Measured 7th harmonic conversion efficiency vs peak power density for the setup up in Fig.S1(a). The fit follows the 5[th] power of the peak power density. The field is tuned off resonance to 2150nm, is incident at 30°, with 20Hz repetition rate.

We now focus on the geometrical resonance centered near 1400nm and scan the carrier wavelength of the pump field across it. The result is plotted in Fig.4, where we show measurements and simulations of the THG conversion efficiency as a function of wavelength, as well as the dispersive character of $\tilde{\beta}$ across the resonance – Fig.4b. Fig.4(a) contains two simulated curves, one obtained with a constant $\tilde{\beta}$ at all wavelengths, and one calculated using the dispersive coefficient $\tilde{\beta}_{3\omega}(\lambda)$ shown in Fig.4(b). While a constant $\tilde{\beta}$ yields good qualitative agreement across the resonance, it is only when using our full dispersive model that qualitative and quantitative agreement can be achieved. This is clearly shown in Fig. 4a by a direct comparison of experiment and the two models with constant and dispersive $\tilde{\beta}_{3\omega}$, respectively. At resonance, the dispersive coefficient $\tilde{\beta}_{3\omega}$ is nearly 1.5 times larger compared to its unitary value off resonance, shown in Fig. 4b. The experimental results clearly point towards a dispersive variation of the model and thus reveal that at least from a classical, macroscopic electrodynamics point of view, theory can be easily reconciled with measurements if we assume that $\tilde{\beta}$ is dispersive across the resonance.

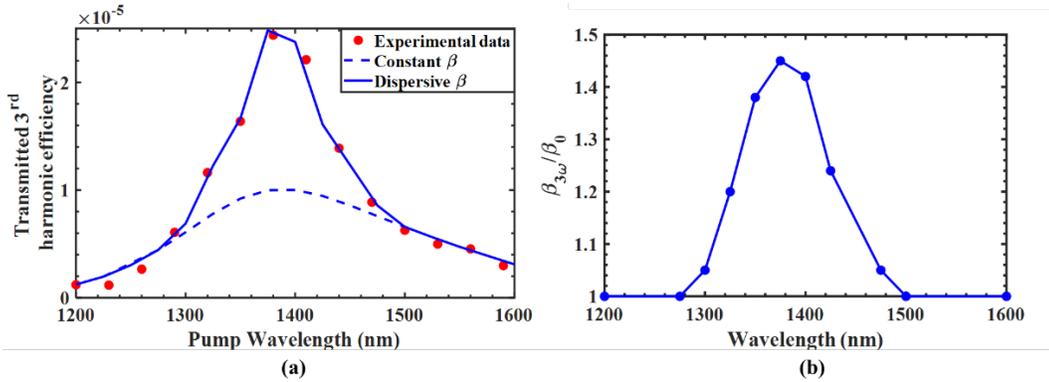

**Fig.4**: (a) Transmitted THG conversion efficiency for constant $\beta$ (dashed blue curve) and with a $\beta_{3\omega}$ that varies as a function of wavelength (solid blue curve) as in Fig.4b. The red markers represent the measured data. (b) Actual functional dependence of the coefficient $\beta_{3\omega}$ used at the third harmonic wavelength. Relevant parameters:
$\tilde{\beta}(3\omega) = \tilde{\beta}_0 = 8.5 \times 10^{-9}$; $\tilde{\delta}(3\omega) = \tilde{\delta}_0 = 4.25 \times 10^{-18}$; $\tilde{\vartheta}(3\omega) = \tilde{\vartheta}_0 = 2.125 \times 10^{-27}$,
$\tilde{\beta}(5\omega) = \tilde{\beta}_0$; $\tilde{\delta}(5\omega) = 9\tilde{\delta}_0$; $\tilde{\vartheta}(5\omega) = 5\tilde{\vartheta}_0$, $\tilde{\beta}(7\omega) = 4\tilde{\beta}_0$; $\tilde{\delta}(7\omega) = \tilde{\delta}_0$; $\tilde{\vartheta}(7\omega) = \tilde{\vartheta}_0$.

In Fig.5(a) we show 4[th] harmonic generation conversion efficiency as a function of incident peak power density for fixed angle. In Fig.5(b) we fix the incident peak power density and perform an angular scan to reveal the angular dependence of the 4[th] harmonic. Once again, the agreement between experiment and theory is remarkable and, to the best of our knowledge, unprecedented. The only free parameters are the effective free electron masses at the different harmonic wavelengths, and the bulk coefficients in Eq.2. SHG results for similar samples were previously reported [47], while the 6th harmonic was not detectable for the intensities we used in this setup (up to 100GW/cm²).

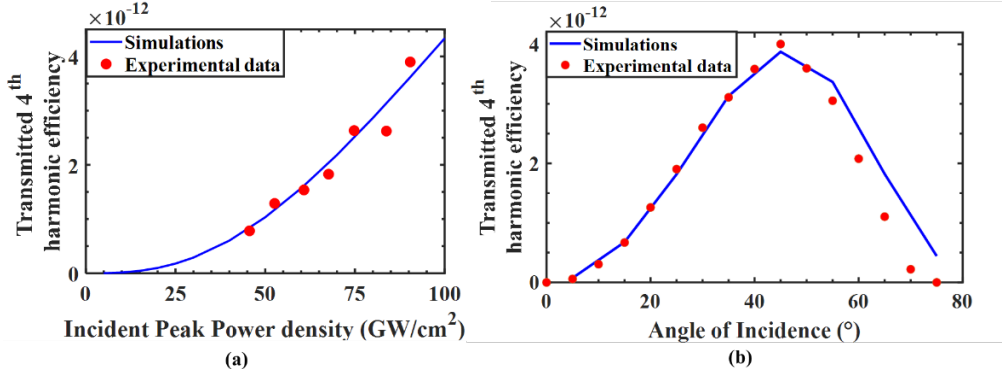

**Fig.5**: (a) Transmitted 4th harmonic generation conversion efficiency vs. incident peak power density with angle of incidence fixed at 30º, carrier wavelength of 1550nm and repetition rate of 20Hz. The red solid markers denote the data, while the simulations are reported as a solid blue curve. (b) Measured and simulated angular dependence of the 4th harmonic signal as a function of incident angle for fixed incident peak power density of 75GW/cm². Data are shown with red markers and simulations are shown as blue solid curves. Relevant parameters: $m_b^*(2\omega) = 0.24 m_e; m_b^*(4\omega) = 0.002 m_e; \tilde{\beta} = 1.5\times10^{-9}; \tilde{\delta} = 2.25\times10^{-18}; \tilde{\vartheta} = 2.25\times10^{-29}$

In Fig.6(a) we plot transmitted THG conversion efficiency vs incident peak power density for the 85fs pulse tuned to 1475nm. Both measured data and simulation display a maximum near 2TW/cm², followed by a decrease in efficiency. We refer to Figs. 6(b) and (c) to explain this behavior. In Fig.6(b) we simulate and plot the effective dielectric constant as a function of time as the pulse traverses the thin film, for a peak power density of approximately 1TW/cm². The simulation reveals a maximum change of order $Re(\delta\epsilon) \sim -4$ when the peak of the pulse reaches the center of the layer. The extraction of this parameter and the method is described in detail in reference [13] and requires knowledge of the constitutive relations. Suffice it to say here that in essence we perform linear and nonlinear numerical ellipsometry on the sample, by calculating the macroscopic dielectric constant as $<\varepsilon(t)> = 1 + 4\pi \frac{<P(t)>}{<E(t)>}$, where the brackets stand for spatial average inside the layer at each instant of time. This definition coincides with the result of an experimental ellipsometric measurement, as demonstrated in reference [13]. Then, the complex change of the dielectric constant may be calculated as the difference between the high and the low intensity dielectric functions as follows: $<\delta\varepsilon> = <\varepsilon_{NL}> - <\varepsilon_L> = 4\pi\left[\frac{<P_{NL}>}{<E_{NL}>} - \frac{<P_L>}{<E_L>}\right]$, where the subscript *L* and *NL* stand for linear and nonlinear, respectively. A dynamic reduction of the dielectric constant from approximately 13 to 9 at the pump wavelength causes an intensity dependent blueshift of the resonance, shown in Fig.6c, effectively detuning the pump out of resonance and reducing the conversion efficiency.

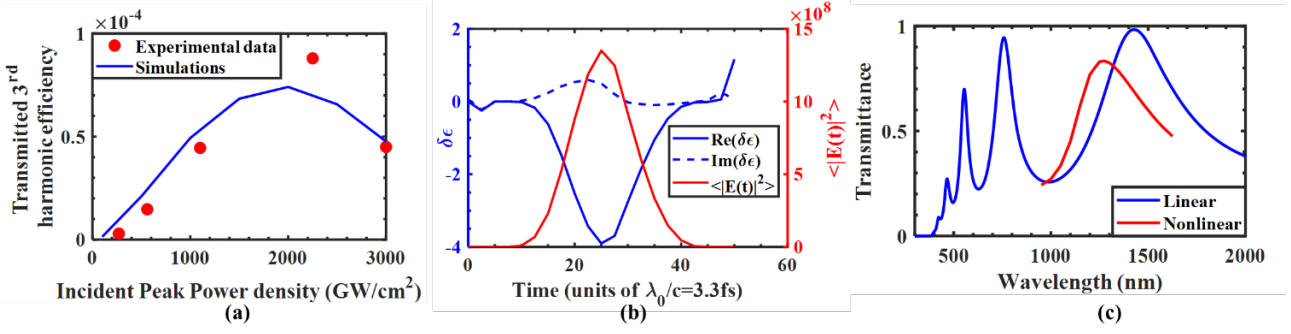

**Fig.6**: (a) Measured and simulated transmitted THG efficiency at high intensities. Pump is tuned at 1475nm with a pulse duration of 85fs; (b) Simulated extraction of the effective dielectric constant from the constitutive relations following the procedure developed in reference [13], for peak power densities of order 1TW/cm². (c) Linear (blue curve) and nonlinear (red curve) transmission functions showing the blueshift of the resonance for a 200nm thick etalon. The relevant parameters are:
$\tilde{\beta}(\omega) = 5\times10^{-10}; \tilde{\beta}(3\omega) = 2\times10^{-10}; \tilde{\beta}(5\omega) = 5\times10^{-11}; \tilde{\beta}(7\omega) = 5\times10^{-11}$,
$\tilde{\delta}(\omega) = 1.25\times10^{-19}; \tilde{\delta}(3\omega) = 5\times10^{-20}; \tilde{\delta}(5\omega) = 1.25\times10^{-20}; \tilde{\delta}(7\omega) = 1.25\times10^{-20}$; ,
$\tilde{\vartheta}(\omega) = 6.25\times10^{-29}; \tilde{\vartheta}(3\omega) = 2.5\times10^{-29}; \tilde{\vartheta}(5\omega) = 6.25\times10^{-30}; \tilde{\vartheta}(7\omega) = 6.25\times10^{-30}$; ,
$k_f(\omega) = 1.6\times10^{-9}; k_f(3\omega) = 6.4\times10^{-10}; k_f(5\omega) = 1.6\times10^{-10}; k_f(7\omega) = 1.6\times10^{-10}$..

At this point a note about parameter space is in order. The total nonlinear bulk polarization that is inserted into Maxwell's equation is the sum of the combined solutions of Eqs.1 and 3, while the instantaneous nonlinear potential may be thought of as a combination of free and bound polarizations as follows:

$$\mathbf{P}_{NL,j} = \kappa_f\left(\mathbf{E}\cdot\mathbf{E}\right)\mathbf{E} + \tilde{\beta}_{bj}(\mathbf{P}_{bj}\cdot\mathbf{P}_{bj})\mathbf{P}_{bj} - \tilde{\delta}_{bj}(\mathbf{P}_{bj}\cdot\mathbf{P}_{bj})^2\mathbf{P}_{bj} + \tilde{\theta}_{bj}(\mathbf{P}_{bj}\cdot\mathbf{P}_{bj})^3\mathbf{P}_{bj} \qquad (4)$$

When the fields and polarizations are expanded up to the 7th harmonic frequency, each term in Eq.(4) contributes to each of the harmonics, including even harmonics. For instance, some of the terms that oscillate at the fundamental wavelength are proportional to $\kappa_f|E_\omega|^2 E_\omega$, $\tilde{\beta}_j|P_\omega|^2 P_\omega$, $\tilde{\delta}_{bj}|P_\omega|^4 P_\omega$, and $\tilde{\theta}_{bj}|P_\omega|^6 P_\omega$. Similarly, just a few of the terms that oscillate at the second harmonic are $\kappa_f|E_\omega|^2 E_{2\omega}$ $\tilde{\beta}_j|P_\omega|^2 P_{2\omega}$, $\tilde{\delta}_{bj}P_\omega^4 P_{2\omega}^*$, and $\tilde{\vartheta}_{bj}|P_\omega|^4 P_{2\omega}^*$. At the third harmonic

wavelength some of the terms are proportional to $\kappa_f E_\omega^3$, $\tilde{\beta}_j P_\omega^3$, $\tilde{\delta}_{bj}|P_\omega|^2 P_\omega^3$, $\tilde{\theta}_{bj}|P_\omega|^4 P_\omega^3$. The leading cubic terms can balance each other, while 5th and 7th powers of the fields tend to either quench or reinforce terms of lower order, depending on the magnitude of the coefficients, as well as material dispersion, since the polarizations intrinsically contain the susceptibilities, which may be either positive or negative depending on tuning. Therefore, the set of parameters that we use in our simulations can be modified somewhat and still yield similar results.

In summary, we have reported odd and even nonlinear frequency conversion up to the 7th harmonic from a subwavelength silicon film. Several of the harmonics are generated in the opacity range, where the material displays metallic behavior with a negative dielectric constant. Those harmonics are transmitted thanks to a phase locking mechanism that allows the inhomogeneous component to effectively defeat nonlinear absorption and allows the harmonic to resonate and propagate through thick layers, if the pump is tuned in the transparency range. Our measurements and simulations show that harmonic generation deep in the UV range is possible, can be efficient under resonant conditions [19, 47, and Supplemental Information], and the sample can withstand peak power densities well above 1TW/cm$^2$. It is remarkable that the reported conversion efficiencies are quite high deep in the UV range, where the material is metallic and absorptive. It is notable that the efficiencies observed for silicon membranes rival the conversion efficiencies reported in transparent conductive oxides pumped near the ENZ crossover wavelength [59]. However, here we achieve the results by exploiting a simple resonant structure and more advanced fabrication processes [60].

**Acknowledgments**

M.A.V. and D.de C. acknowledge partial funding from NATO SPS Grant no. G5984. M.A.V. thanks MUR as part of the PRIN 2022 project PILLARS (2022YJ5AZH). M.F. acknowledges financial support from EPSRC project ID: EP/X035158/1 and AFOSR (EOARD) under Award No. FA8655-23-1-7254. S.M., J.T., and C.C. acknowledge US Army Research Laboratory Cooperative Agreement N° W911NF-22-2-0236 issued by US ARMY ACC-APG-RTP. Y.K. was supported by the Australian Research Council (grant DP210101292) and the International Technology Center Indo-Pacific (ITC-IPAC) via Army Research Office (contract FA520923C0023). The authors thank R. Davidson, P. Tonkaev, M. Clerici and N. Litchinitser for critical reading of the manuscript.**References**

# Supplemental Information:

# High-harmonic generation from subwavelength silicon films

K. Hallman[1], S. Stengel[2], W. Jaffray[2], F. Belli[2], M. Ferrera[2], M.A. Vincenti[3], D. de Ceglia[3], Y. Kivshar[4], N. Akozbek[5], S. Mukhopadhyay[6], J. Trull[6], C. Cojocaru[6], and M. Scalora[7*]

[1]*PeopleTec, Inc. 4901-I Corporate Dr., Huntsville, AL 35805, USA*
[2]*Institute of Photonics and Quantum Sciences Heriot-Watt University, SUPA Edinburgh, EH14 4AS United Kingdom*
[3]*Department of Information Engineering – University of Brescia, 25123 Brescia, Italy*
[4]*Nonlinear Physics Centre, Australian National University, Canberra, ACT 2601, Australia*
[5]*US Army Space & Missile Defense Command, Tech Center, Redstone Arsenal, AL 35898 USA*
[6]*Department of Physics, Universitat Politècnica de Catalunya, 08222 Terrassa (Barcelona), Spain*
[7]*FCDD-AMT-MGR, DEVCOM AvMC, Charles M. Bowden Research Center, Redstone Arsenal, Alabama, 35898-5000, USA*

*\*michael.scalora.civ@army.mil*

## A. Experimental Setups

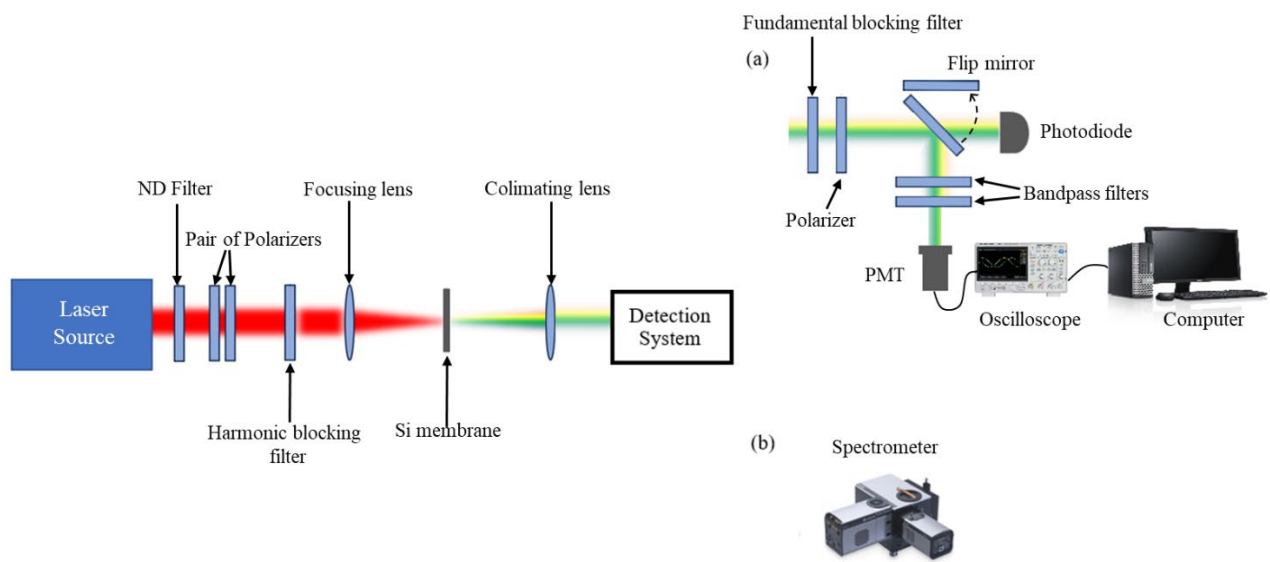

**Fig.S1:** Experimental setup used for measurement of conversion efficiencies as a function of incident pump power and angle for (a) peak power densities below 100GW/cm$^2$; and (b) used for the measurement of the spectral response at peak power densities above 100GW/cm$^2$.

Two experimental setups were deployed to measure different aspects of harmonic generation and are shown schematically in Fig.S1. The first setup (Fig.S1a) is used for peak power densities below 100GW/cm$^2$ and is like the one outlined in Ref. [47] of the manuscript. This system is used to measure conversion efficiencies as a function of peak power density and angular dependence of the harmonic signals. It consists of an optical parametric amplifier (Coherent Opera Solo) pumped by an amplified Ti:Sapphire laser system (Coherent Astrella) and used as a source of femtosecond pulses with tunable wavelength and repetition rate. To prepare the pump beam, the OPA output was carefully filtered to allow through only the desired pump wavelength; neutral density filters and a pair of crossed polarizers adjusted the pump power, and a calcium fluoride lens with a long focal length was used to focus onto the samples. A second calcium fluoride lens collected the harmonic beam, and a polarizer was used to select the harmonic polarization. A mirror on a flip mount allowed the selection of either a calibrated silicon photodiode, or a more sensitive photomultiplier tube (PMT) to enable calibration of the responsivity of the full detection system including electronics used to collect the signal. The samples were mounted on a custom goniometer with six degrees of freedom, and the angle of incidence could be adjusted using a pair of rotational stages after aligning the sample to the center of rotation. For each point in the pump wavelength sweeps, the pump wavelength was changed manually, the laser energy adjusted to compensate for the wavelength dependent OPA efficiency, a separate responsivity value was used for the detection system, and a factor was used to compensate for the wavelength-dependent transmission (or reflection) of all optics. To measure higher order harmonics, especially the weaker even order harmonics, the intensity of the pump needed to be increased. Thermal loading of the samples was avoided by decreasing the repetition rate, sometimes as low as 2 Hz, and the integration time was increased.

Signal detection in the 200nm range was performed in a different lab with a different setup and laser system (shown in Fig.S1b). A Ti:Sapphire laser was used in conjunction with an OPA to generate 85fs pulses centered at 1475nm with a repetition rate of 10Hz. After the OPA, a visible filter was used to remove any residual pump with wavelengths shorter than 800nm. The beam is p-polarized, attenuated to the required power, and focused onto the film at normal incidence through a 200mm focal-length lens. This results in a laser spot size on the sample of 350μm (FWHM). After generating the higher harmonics, an uncoated $CaF_2$ lens was used to collimate the beam. A grating was then used to spatially separate the harmonics, which are then sent through a slit and recorded by a photodiode (an integrated spectrometer was used to verify that the free space grating spectrometer was calibrated correctly). An input power of 1.5mW was used, which corresponds to a peak intensity of 3TW/cm$^2$ on the sample. The sensitivity of the detection system was such that this peak power density was necessary to detect the 7$^{th}$ harmonic at 210nm, which is the lower limit of our detection system, but not the lower limit of what is possible.

## B. 7$^{th}$ harmonic generation vs incident wavelength.

We simulated HHG across the spectrum shown in Fig.1b of the manuscript. The figure shows that it is possible to achieve relatively efficient 7$^{th}$ harmonic generation below 100nm despite the presence of absorption at the pump wavelength, and despite the absence of local field amplification (see item **C.** below.) These results suggest that efficient harmonic generation below 100nm is possible by appropriately sizing a nanowire array made of materials like Si or GaP (see reference [19] in the manuscript).

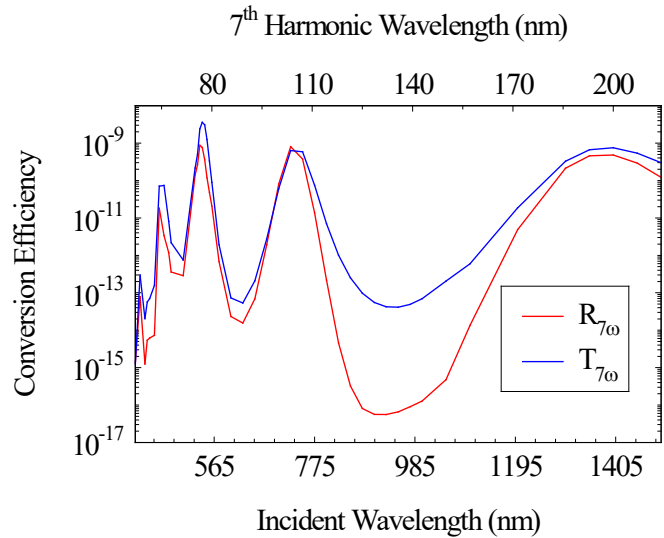

**Fig.S2:** Simulated reflected (R) and transmitted (T) 7$^{th}$ harmonic conversion efficiency vs incident pump wavelength in a range that spans the spectrum shown in Fig.1b., and incident peak power densities of 100GW/cm$^2$. All things being equal, the resonance near 532nm is predicted to yield conversion efficiencies nearly one order of magnitude larger compared to the 1400nm resonance, despite partial absorption of the pump beam.

## C. Fabry-Perot (FP) etalon.

In the simple FP etalon we are considering, the **H** and **E** fields delocalize, but the local electric field intensity is *not* amplified, notwithstanding the fact that the energy velocity of the pulse slows down to approximately $V_e=0.2c$. This slow down corresponds to multiple passes that give rise to field maxima and minima inside the layer, with magnetic and electric field intensities tuned to 1400nm and 756nm as indicated in the figure.

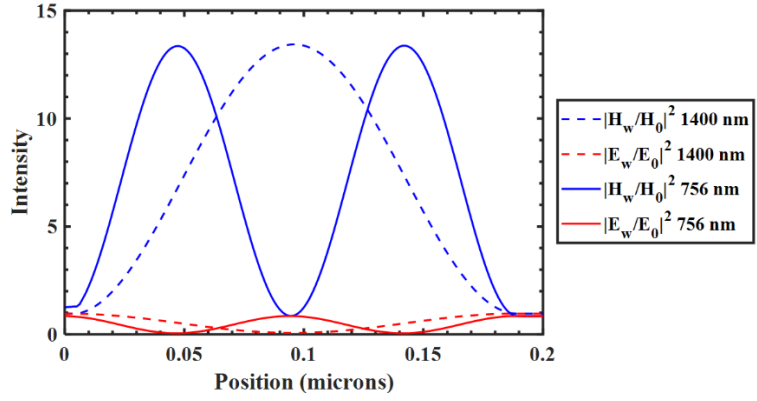

**Fig.S3:** Fabry-Perot modes at 1400nm and 756nm. The electric field intensity is not amplified. In each case, an H-field maximum coincides with an E-field minimum, leading to a reduction of the Poynting vector and decreased energy velocity.